\documentclass[journal]{IEEEtran}
\IEEEoverridecommandlockouts

\usepackage{multirow}
\usepackage{cite}

\ifCLASSINFOpdf
\else
\fi
\usepackage{graphicx}
\graphicspath{{figures/}} 
\usepackage[utf8]{inputenc}

\usepackage[cmex10]{amsmath}
\usepackage{amssymb}

\renewcommand{\baselinestretch}{0.96}

\usepackage{amsmath}

\DeclareSymbolFont{matha}{OML}{txmi}{b}{it}
\DeclareMathSymbol{\varv}{\mathord}{matha}{118}

\begin{document}

%
\title{Zero-inertia Systems: Sufficient Conditions for Phasor Modeling 
}
\author{George~S.~Misyris,~\IEEEmembership{Student Member,~IEEE,}
        Spyros~Chatzivasileiadis,~\IEEEmembership{Senior Member,~IEEE,}
        and~Tilman~Weckesser,~\IEEEmembership{Member,~IEEE} \vspace{-16pt}
}
\maketitle


\begin{abstract}
Time-domain simulations are a critical tool for power system operators. Depending on the instability mechanism under consideration and the system characteristics, such as the time constants of controllers, either phasor or Electro-Magnetic Transient (EMT) models should be employed. On the one hand, EMT models provide a detailed-modeling of the system dynamics, thus increase the reliability of stability analysis; on the other end, using these models increase the computational times of simulations, slowing down the security assessment process. To decrease computational time, system operators could resort to phasor-mode simulations for a (hopefully large) subset of disturbances. This paper investigates the appropriateness of phasor-approximation models on simulating events related to power supply and balance stability in zero-inertia systems. First, the stability boundaries, which each model is able to identify, are analyzed; then sufficient conditions for control parameters are derived, which allow using phasor-approximation models to monitor power sharing among grid-forming converter-based resources. Time-domain simulations are performed in PowerFactory DigSilent to verify the results.

\end{abstract}

\vspace{-14pt}
\section{Introduction}
\vspace{-2pt}
\IEEEPARstart{T}{he} replacement of synchronous generators with converter-based resources is followed by a decrease of system inertia \cite{Milano2018,markovic2019,Collados2019,Misyris2018,Misyris2019}. When the replacement is only partial, the system is referred to as low-inertia power system. In case of complete absence of synchronous generation, it becomes a zero-inertia system. 
In zero-inertia systems, the characteristic time-constants of converter dynamics are much shorter compared to synchronous generators. This is due the fact that the converters need to act fast in order to maintain the power balance \cite{Farrokhabadi2019}. Thus, there is no distinct separation between converter control loops and eigen-frequencies in power network \cite{markovic2019}. As a result, the influence that electro-magnetic dynamics have on system stability is growing and phasor-approximation modeling of transmission lines \cite{kundur,Turhan2008} becomes inappropriate for determining the stability boundaries of the system.
Consequently, to analyze fast power components, such as Power Electronic (PE) devices, the electro-magnetic dynamics (network line dynamics) are vital for system stability. However, taking electro-magnetic dynamics into account drastically increases the computational time for solving the system of differential algebraic equations, which describes the power system dynamics (since the number of state variables increases). 

Several works have focused on reduced-order modeling of AC grids with penetration of converter-based generation units \cite{Vorobev2018,Gu2018,Purba2019,Purba2019b}, to decrease the computational time of EMT simulations. In \cite{Vorobev2018}, a high-fidelity model order reduction is proposed for proper modeling of converter-based micro-grids. The authors formulate a general method for stability analysis by using a first-order approximation of the singular perturbation theory, as opposed to zero-order which corresponds to neglecting the dynamics of fast variables. By employing this method they are able to capture the influence of ``fast" dynamics on the stability of slower modes. In \cite{Gu2018}, a reduced-order model for representing converters was studied, where the authors proposed a method in which several fast states are disregarded but a representation of their interaction with the slow states is taken into account. Aggregate dynamic models of grid-tied three phase converters has been also presented in \cite{Purba2019} and \cite{Purba2019b}. 

While most of the literature focuses on reduced-order modeling of power converters, to the best of our knowledge, there has been no work to derive sufficient conditions for using  phasor-approximation modeling for simulating zero-inertia systems. In zero-inertia systems with multiple grid-forming converters, power sharing is achieved by proper tuning of their control parameters and in particular of their frequency droop-based active power controllers. Unlike conventional power grids, tuning of frequency droops (in order to effectively determine the power sharing) may trigger certain type of instability phenomena, such as voltage and harmonic problems, which are not captured by RMS modeling. In our work, we are interested in assessing the limitations of RMS modeling for simulating active power sharing in zero-inertia systems. 

Hence, the scope of this paper is to identify the appropriateness of phasor modeling in zero-inertia systems, considering the control parameters of grid-forming Voltage Source Converters (VSCs) \cite{Rocabert2012}. Our goal is to assess under which conditions phasor-approximation modeling (used by simulation tools, such as PowerFactory DigSilent, PSSE, etc.) can still be used for performing power system dynamic security assessment in zero-inertia systems \cite{Kundur2003}. The motivation is that system operators could continue using the phasor approximation for monitoring the active power sharing in their systems even during zero- or low- inertia periods, as long as the mismatch between the phasor-approximation and the EMT models are within acceptable limits. 

The contributions of this paper are the following:
\begin{enumerate}
    
    \item We derive sufficient stability rules for zero-inertia systems both in case of RMS and EMT modeling, in order to validate under which conditions RMS modeling can lead to wrong conclusions regarding the system stability. 
    \item We highlight how the control parameters of the grid-forming VSCs influence the small-perturbation stability of the system. 
    \item We propose a robust control design of the grid-forming converters that allows for sufficient damping of critical eigen-frequencies in power network, so that RMS modeling can still be used for monitoring/simulating the power sharing in a zero-inertia system.
\end{enumerate}


The rest of the paper is organized as follows. In Section II, we present a description of the dominant-dynamics of a zero-inertia system. In section III, we derive sufficient conditions regarding the small-signal stability analysis of the RMS and EMT models. In section IV, we first identify the differences in the control design when phasor-approximation modeling is used. Then we propose a methodology for analyzing the performance of the RMS model based on frequency domain specifications. Section V verifies the conditions derived in section III using time-domain simulations. Conclusions are drawn in \mbox{Section VI}.
\vspace{-8pt}
\section{System Modeling}
\vspace{-2pt}
\subsection{Zero-inertia systems}
Zero-inertia systems consist of 100 \% converter-based generation units. We can find these systems both in distribution and offshore transmission level \cite{Vorobev2018,Raza2018,Pogaku2007,SIMPSONPORCO20132603}. The frequency and voltage of such a system are set by PE devices known as grid-forming converters \cite{Rocabert2012,Farrokhabadi2019}. To assess the accuracy of phasor-approximation models for simulating zero-inertia systems, we need first to understand the operating principles of grid-forming converters and the control system stability issues that may arise due to inadequate control schemes (as it is the main element that determine the power supply and balance stability \cite{Farrokhabadi2019}). Hence, in this section we present the control structure and the dynamics describing the operation of a grid-forming converter. 
\vspace{-8pt}
%
\subsection{Grid-forming converter} \label{Gridform}
As described in \cite{Rocabert2012}, grid-forming converters, equipped with the controller illustrated in Fig. \ref{fig:GridformingConv}, can be represented as ideal AC voltage sources which set the voltage amplitude V and frequency w of the local grid.
\subsubsection{$dq$ and $xy$ frames}
Similar to \cite{Papangelis2018}, we define two rotating frames: i) the $dq$ rotating frame and ii) the $xy$ rotating frame. The angular speed of the $dq$ frame is equal to the frequency of the $i^{th}$ converter $\omega_i$, whereas the angular speed of the $xy$ frame is equal to the frequency of the grid $\omega_g$. Similar to \cite{Harnefors2018}, the converter $dq$ frame leads the $xy$ frame by the angle $\Delta \theta$, which is given by: 
\begin{equation}
    \centering
    \frac{d\Delta\theta}{dt} = (\omega_i-\omega_g)\omega_b 
\end{equation}
where $\omega_b = 100 \pi $ $rad/s$ is the base angular speed. The transformation from the $xy$ to $dq$ coordinates is given by:
\begin{equation}
    \boldsymbol{x} = \boldsymbol{{x}^c}e^{-j(\theta_0+\Delta\theta)}
\end{equation}
where the superscript $c$ denotes the $xy$ frame. The corresponding $dq$ frame is denoted without a superscript, e.g., $x$. The quantities in $dq$- and $xy$ frame are described in complex space vector form as:
\begin{subequations}
\begin{align}
    \boldsymbol{x} &= x^d+jx^q\\ 
    \boldsymbol{x}^{c} &= x^x+jx^y.
\end{align}
\end{subequations}
\subsubsection{Phase reactor and transformer}
\begin{figure}[t]
    \centering
        \includegraphics[scale=1.0]{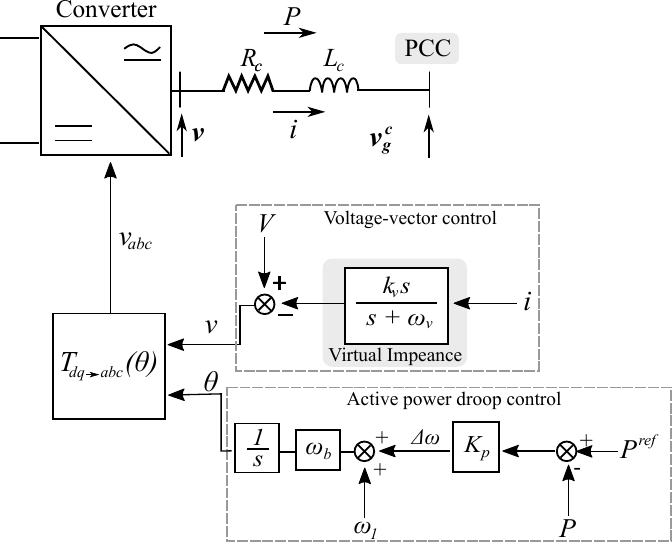}
    \vspace{-5pt}
    \caption{Basic control structure of converter operating as a voltage source.}
    \label{fig:GridformingConv}
    \vspace{-10pt}
\end{figure}

As depicted in Fig. \ref{fig:GridformingConv}, the circuit model consists of a VSC connected through an $RL$ impedance to the rest of the grid. $L_c$ represents the sum of the phase reactor and transformer inductances. $R_c$ represents the sum of the phase reactor and transformer resistances. We remind here that phase reactors are $RL$ filters used for reducing the voltage harmonics caused by the Pulse-Width Modulation (PWM). In case of a modular multilevel converter, the phase reactor can be neglected. Thus the $RL$ circuit consists of the resistance and inductance of the transformer. The relation between the current flowing through the $RL$ circuit and the voltages $\boldsymbol{\varv}$ and $\boldsymbol{{\varv}}_g$, expressed in the $dq$ frame, is given by:
\begin{equation}
    \centering
    \boldsymbol{\varv} = \boldsymbol{{\varv}}_g^c e^{-j(\theta_0+\Delta\theta)} + [R_c+(s+j\omega_i)L_c]\boldsymbol{i}
    \label{Eq:phasereactor}
\end{equation}
where $L_c$ and $R_c$ are expressed in p.u. and $s$ is the Laplace complex variable.
$\boldsymbol{{\varv}}_g^c$ is the grid voltage expressed in the $xy$ frame and $\boldsymbol{\varv}$ is the converter terminal voltage in the $dq$ frame. 

\subsubsection{Voltage-vector control}
A voltage set-point is used to control the output voltage $\boldsymbol{\varv}$ of the offshore converter at its terminal bus. 
Apart from setting the voltage at the converter terminal, the voltage-vector control unit must be able to provide damping of poorly-damped resonances. To this end, the virtual impedance concept is incorporated to increase the damping of high frequency modes. As indicated by its name, the virtual impedance acts as an additional $RL$ circuit connecting the converter with the Point of Common Coupling (PCC), as shown in \eqref{Eq:VirImp}. 
\begin{equation}
    \centering
    \boldsymbol{\varv} = V_{\rm set}- [\underbrace{r_{tv}+j\omega_il_{tv}}_{Z(s)}]\boldsymbol{i}
    \label{Eq:VirImp}
\end{equation}
where $\footnotesize{Z(s)=r_{tv}+j\omega_il_{tv}}$ denotes the virtual impedance, with $r_{tv}$ and $l_{tv}$ being the tuning parameters. 
To realize the virtual impedance control scheme, we substitute $Z(s)$ \eqref{Eq:VirImp} with a high pass filter \cite{ZhangThesis}. By using a high pass filter, we can eliminate the voltage static error and react only during transients. To this end, the voltage-vector control can be expressed as:
\begin{equation}
    \centering
    \boldsymbol{\varv} = V_{\rm set}- \underbrace{\frac{k_{\varv} s}{s+\omega_{\varv}}}_{H_{\rm HP}(s)}\boldsymbol{i}
\end{equation}
where $k_{v}$ determines the damping effect on the poorly-damped high frequency modes, caused by the multiple cables. The cut-off frequency of the high-pass filter ($H_{\rm HP}(s)$) is determined by the value of $\omega_{\varv}$, which according to \cite{Harnefors2018} takes values in the range 0.1$\omega_1\rightarrow$ 0.2$\omega_1$, where $\omega_1$ is the reference angular speed chosen to be 1.0 p.u. 
\subsubsection{Active power droop control}
The active power droop control regulates the angular speed $\omega_i$; the controller acts by adjusting the voltage angle $\theta_i$ based on deviations of the injected/absorbed active power $P$ with respect to a given reference value $P_{\rm ref}$.
The state-space equations describing the imposed frequency and angle at each converter terminal are given by:
\begin{subequations}
\begin{align}
\vspace{-10pt}
    \omega_i &= \omega_{\rm 1}+K_p(P_{\rm ref}-P) \label{Eq::2a} \\
    \frac{d\theta_i}{dt}&=\omega_i\omega_b
    \vspace{-10pt}
\end{align}
\end{subequations}
where the frequency droop $K_p$ is used to regulate the frequency $\omega_i$ based on the power mismatch between the power reference point and the power output.
\vspace{-2pt}
\vspace{-4pt}

\section{Stability analysis using RMS and EMT model}
The aim of this section is to derive sufficient stability rules both in case of RMS and EMT modeling, in order to validate under which conditions RMS modeling can lead to wrong conclusions regarding the system stability. Moreover, we aim at analyzing how the control parameters of the grid-forming VSCs influence the small-perturbation stability of the system. 

To show this, we investigate the stability of the system depicted in Fig. \ref{fig:GridformingConv}, both in case of RMS and EMT modeling. The system consists of only one grid-forming converter connected through a phase reactor/ transformer to the grid voltage $\boldsymbol{\varv_g^c}$. For the analysis, we assume that the magnitude of the grid voltage $\boldsymbol{\varv_g^c}$ remains constant for small disturbances. This assumption holds in power networks with multiple grid-forming converters and as far as AC faults are not considered. This is due to their robust performance in the presence of low-short circuit ratio, which results in a limited voltage drop \cite{Zhang2010}. Thus the value of the converter's reference voltage $V$ is kept almost constant (we do not consider a $QV$ characteristic for regulating the reactive power). To this end, we do not investigate the transfer function between the reactive power and the voltage ($\Delta Q$ versus $\Delta V$).

\begin{figure}[t!]
    \centering
                \includegraphics[scale=2]{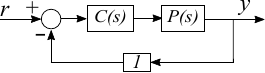}
        \vspace{-4pt}
    \caption{Closed loop system: $C(s)$ is the controller and $P(s)$ is the system plant.}
    \label{fig:OpenLoopSystem}
            \vspace{-16pt}
\end{figure}
As mentioned in the introduction, 
we aim at identifying, for each model (EMT and RMS), small-perturbation stability issues \cite{Farrokhabadi2019} which may occur due to poor tuning of grid-forming converters.
Hence, we focus on the transfer function describing the dynamics of the active power controller. As an intermediate step, we derive the transfer function of $\Delta P(s)$ to $\Delta \theta(s)$ (\mbox{$G_{\theta P}(s)=\frac{\Delta P(s)}{\Delta \theta(s)}$}). 
In case of EMT modeling, we refer the reader to \cite{Harnefors2018} for the derivation of the transfer function. As for the RMS model, the derivation of the transfer function is presented in the appendix of this paper. 

Having derived $G_{\theta P}(s)$, we can obtain the transfer function of the closed loop system describing the dynamics of the active power controller. Consider the system in Fig. \ref{fig:OpenLoopSystem}, where $C(s)=\frac{K_p}{s}$ is the frequency droop controller and $P(s)=G_{\theta P}(s)$ is the plant. The relation $\Delta P$ to $\Delta P_{\rm ref}$ is given by:
\begin{equation}
\small
    G_{P {\rm cl}}(s)=\frac{\Delta P(s)}{\Delta P_{\rm ref}(s)}=\frac{\frac{K_p}{s}G_{\theta P}(s)}{1+\frac{K_p}{s}G_{\theta P}(s)}.
\end{equation}
In the rest of the paper, the resistance of the RL circuit connected between the converter terminals and the PCC is neglected. As a result, $R_c \ll L_c$ (unless otherwise stated).

\vspace{-3pt}
\subsection{EMT model of closed-loop system}
To analyze the stability of the closed loop system, in case of EMT modeling, we employ the transfer function \eqref{Eq:GpiEMT} (derived in \cite{Harnefors2018}), which describes the relation between the angle \mbox{$\Delta\theta$ and $\Delta P$} :
\makeatletter
    \def\tagform@#1{\maketag@@@{\normalsize(#1)\@@italiccorr}}
\makeatother
\begin{equation}
\footnotesize{
    \label{Eq:GpiEMT}
    G_{\theta P}^{\rm EMT}(s) =\frac{\Delta P(s)}{\Delta \theta(s)}=\frac{V_{\rm set}^2}{\omega_1 L_c}\cdot\frac{a s^2+(1+a+b(s))\omega_1^2}{s^2+\frac{2H_{\rm HP}(s)}{L_c}s+\omega_1^2+(\frac{H_{\rm HP}(s)}{L_c})^2}}
\end{equation}
where 
\begin{equation*}
    a=\frac{\omega_1 L_c i_{q0}}{V_{\rm set}} \> \> \> \> b(s) = -\frac{H_{\rm HP}^2(s)}{V_{\rm set}}(\frac{i_{q0}}{\omega_1 L_c}+\frac{|i_{0}|^2}{V_{\rm set}}).
    \label{Eq::abPSL}
\end{equation*}
$H_{\rm HP}(s)=\frac{k_{v}s}{s+\omega_{v}}$ is the high pass filter of the virtual impedance control unit. As described in the previous section, the high pass filter is employed to damp eigen-frequencies in power network. In case of EMT modeling, the closed loop transfer function of the active power controller is given by:
\begin{subequations}
\begin{align}
\small
\label{Eq:GpclEMT}
    G_{Pcl}^{\rm EMT}(s) &= \frac{N_{Pcl}^{\rm EMT}(s)}{D_{Pcl}^{\rm EMT}(s)}\\
         \label{Eq:NPclEMT}
    N_{Pcl}^{\rm EMT}(s) &= \frac{K_pV_{\rm set}^2(a s^2 + (1+a+b(s))\omega_{1}^2)}{\omega_1 L_c} \\
        \vspace{4pt}
        \label{Eq:DpclEMT}
    D_{Pcl}^{\rm EMT}(s) &= s^3+(2 \frac{H_{\rm HP}(s)}{L_c}+\frac{V_{\rm set}^2 K_p a}{\omega_1 L_c})s^2  \\\vspace{4pt}
    & + s(\omega_1^2+ (\frac{H_{\rm HP}(s)}{L_c})^2)+\frac{K_p V_{\rm set}^2\omega_1}{ L_c}(1+a+b(s)). \nonumber
\end{align}
\end{subequations}

\vspace{-10pt}
\subsection{RMS model of closed-loop system}
The RMS model neglects the electro-magnetic transients. Thus, the relationship between the power and angle can be described: 
\begin{equation}
\footnotesize
\label{Eq:GthetaRMS}
    G_{\theta P}^{\rm RMS}(s) =\frac{\Delta P(s)}{\Delta \theta(s)}= \frac{V_{\rm set}^2}{L_c}\frac{(1+a+b(s))\omega_1}{\omega_1^2+(\frac{H_{\rm HP}(s)}{L_c})^2}.
\end{equation}
In case of RMS modeling, the closed loop transfer function of the active power controller is given by:
\begin{subequations}
\small
\begin{align}
\centering
\label{Eq:GpclRMS}
     G_{Pcl}^{\rm RMS}(s) &= \frac{N_{Pcl}^{\rm RMS}(s)}{D_{Pcl}^{\rm RMS}(s)} \\
     \label{Eq:NPclRMS}
     N_{Pcl}^{\rm RMS}(s) &= \frac{K_pV_{\rm set}^2(1+a+b(s))\omega_{1}}{L_c} \\
    \vspace{2pt}
         \label{Eq:DPclRMS}
    D_{Pcl}^{\rm RMS}(s) &= s(\omega_1^2+(\frac{H_{\rm HP}(s)}{L_c})^2)+\frac{K_p V_{\rm set}^2 \omega_1}{L_c}(1+a+b(s)).
\end{align}
\vspace{-20pt}
\end{subequations}
\subsection{Impact of virtual impedance on small-signal stability} \label{SectionIIIC}
The virtual impedance plays an important role on increasing the damping ratio of poorly-damped eigen-frequencies and limiting the oscillations on active power. Considering that the RMS model is unable to eigen-frequencies in power network, we expect a minor impact of the virtual impedance on system small-signal stability. Unlike the RMS model, the EMT one considers network line dynamics. Thus, we expect that the virtual impedance will have a major role on the system stability defined by the EMT model. To show this, we present here two small-signal stability analyses with and without the virtual impedance controller, where we use the Routh-Hurwitz stability criterion in order to identify the stability boundaries of each model. 
\subsubsection{Small-signal stability without virtual impedance}
When the virtual impedance controller is disregarded $H_{\rm HP}(s)=0$, 
and $b(s) = 0$ in \eqref{Eq:GpiEMT} and \eqref{Eq:GthetaRMS}. 
Using the transfer functions derived in the previous subsection, we can make the following observations.
 Looking at the numerator of the open-loop transfer functions, $G_{\theta P}^{\rm EMT}(s)$ and $G_{\theta P}^{\rm RMS}(s)$ in  \eqref{Eq:GpiEMT} and \eqref{Eq:GthetaRMS} respectively, we can see that $G_{\theta P}^{\rm EMT}(s)$ has two zeros (when $a \neq 0$) that can be calculated by solving $s^2=-(\frac{1}{a}+1)\omega_1^2$. If $\frac{1}{a}+1>0$, the zeros lie on the imaginary axis and are equal to $s_{1,2}=\pm j\omega_1\sqrt{\frac{1}{a}+1}$. 
 
 If $\frac{1}{a}+1<0$ instead, the plant has a Right Half Plane (RHP) zero  (nonminimum-phase behavior of the system \cite{Skogestad2005}), which is equal to $s=\omega_1\sqrt{\frac{1}{a}+1}$. A RHP zero acts as a time delay and limits the achievable performance of a feedback controller by limiting the gain margin of the open loop system \cite{Skogestad2005,Zhang2010}. That is not the case when the RMS model is used, since the numerator $N_{Pcl}^{\rm RMS}(s)$ of $G_{\theta P}^{\rm RMS}(s)$ in \eqref{Eq:GthetaRMS} consists only of a static gain, which is proportional to $1+a$. 
 When analyzing the closed-loop transfer functions, it can be noticed that the characteristic polynomials of $G_{Pcl}^{\rm RMS}(s)$ and $G_{Pcl}^{\rm EMT}(s)$ are of different order with different poles. 
 
 In case of RMS modeling, there is one real eigenvalue defining stability of the closed loop system: $s$=$-\frac{K_p V_{\rm set}^2}{\omega_1 L_c}(1+a)$. For any positive value of the frequency droop, the eigenvalue is real and negative as long as $1+a>0$. That implies that the system is asymptotically stable as $t\rightarrow\infty$. 
 
\begin{figure}[t]
    \centering
    \includegraphics[width=\linewidth,trim={3.3cm 11.3cm 3.2cm 11cm},clip]{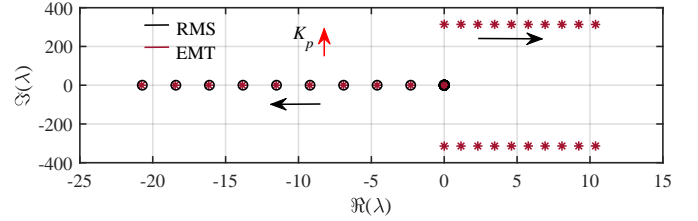}
    \vspace{-15pt}
    \caption{Comparison between RMS and EMT - Without virtual impedance unit.}
    \label{fig:ComparisonAnalyticalWithout}
    \vspace{-18pt}
\end{figure}

 On the contrary, when the EMT model is used, the characteristic polynomial of the closed loop system is of a $3^{rd}$ order. Consider the generic cubic polynomial $\alpha(s) = a_0 s^3 + a_1 s^2 + a_2 s + a_3$ = 0, where
\begin{equation*}
\small
    a_0=1, \> \> \> a_1=\frac{V_{\rm set}^2 K_p a}{\omega_1 L_c}, \> \> \>  a_2=\omega_1^2, \> \> \> a_3=\omega_1^2(\frac{V_{\rm set}^2 K_p}{\omega_1 L_c}(1+a)).
\end{equation*}
To guarantee stability, all roots of the cubic polynomial $\alpha(s)$ must have negative real parts. According to the Routh-Hurwitz stability criterion \cite{rahman2002analytic}, all roots of \eqref{Eq:DpclEMT} will have negative real parts if it holds $a_1a_2 > a_0a_3$. This condition is equivalent to:
\begin{equation}
    0 > \frac{K_p V_{\rm set}^2}{\omega_1 L_c}
    \label{Eq:withoutVi}
\end{equation}
which indicates that the system is unstable for any $K_p > 0 $. 
In case of $K_p = 0$, the poles of the transfer function $G_{Pcl}^{\rm EMT}(s)$ (roots of \eqref{Eq:DpclEMT}) are found to be equal to $s=\pm j\omega_1$. As $K_p$ increases, the eigenvalues move towards the RHP, which verifies \eqref{Eq:withoutVi}. 
Fig. \ref{fig:ComparisonAnalyticalWithout} also verifies our conclusions regarding the asymptotic stability of the system. The figure depicts the root locus when varying $K_p$ in the range of [0, 0.01]. As shown in the figure, for $K_p>0$,  the EMT model indicates that the system is unstable, while the RMS model is unable to predict the instability. It can also be seen that both models have a real eigenvalue in common, which moves towards the Left Half Place (LHP) as $K_p$ increases.
\subsubsection{Small-signal stability with virtual impedance}
For this analysis, the virtual impedance controller is included in the model as a static gain (virtual impedance acts as pure active resistance), hence $H_{\rm HP}(s)=k_{\varv}$ and $\omega_{\varv}\approx0$. This assumption is reasonable as $\omega_{\varv}\rightarrow0$ \cite{Harnefors2018} (the validity of this assumption is also checked in the next Section V through time-domain simulations). As mentioned above, a small value for $\omega_{\varv}$ is necessary if we want to eliminate the voltage static error that is introduced, when a pure active resistance is employed ($H_{\rm HP}(s)=k_{\varv}$). Moreover, since $\omega_{\varv}=0$, $b(s)$ is a static gain \mbox{($b=-\frac{k_{\varv}^2}{V_{\rm set}}(\frac{i_{q0}}{\omega_1L_c}++\frac{|i_{0}|^2}{V_{\rm set}})$)}.
     In case of RMS modeling, there is one real eigenvalue determining the stability of the closed loop system: 
     \begin{equation}
         \centering
         s=-\frac{K_pL_c V_{\rm set}^2\omega_1 (1+a+b)}{k_{\varv}^2+L_c^2\omega_1^2}.
         \label{Eq::Pole1}
     \end{equation}
    According to \eqref{Eq::Pole1}, the system becomes more stable as $K_p$ increases. 
    A useful expression from \eqref{Eq::Pole1} can be obtained for $i_0$ and $i_{q0}$ equal to zero, which yields that the closed loop system is asymptotically stable as $t\rightarrow\infty$ for any positive value of $K_p$.

     In case of EMT modeling, the characteristic polynomial $D_{Pcl}^{\rm EMT}(s)$ of the closed-loop system is of a $3^{rd}$ order. To determine the stability of the system, the Routh-Hurwitz stability criterion is employed, where the coefficients of the generic cubic polynomial (when considering the virtual resistance) are given by:
    \begin{equation*}
        \small
    \begin{split}
    a_0&=1, \> \> \> \> \> \> \> \> \> \> \> \> \> \> \> \> \> \> \> \> \> \> \>  a_1=\frac{V_{\rm set}^2 K_p a}{\omega_1 L_c} + \frac{2 k_{\varv}}{L_c} \> \> \>  \\a_2&= \omega_1^2 + (\frac{k_{\varv}}{L_c})^2,  \> \> \> 
    a_3=\omega_1^2(\frac{V_{\rm set}^2 K_p}{\omega_1 L_c}(1+a+b))
    \end{split}
    \end{equation*}
    As mentioned above, system stability is guaranteed if \mbox{$a_1a_2 > a_0a_3$}. The condition needed to be satisfied is:
    \begin{align}
    \small
    \centering
      &  (\frac{{k_{\varv}}^2}{L_c^2}+{\omega_1}^2) ({i_{q0}} {K_p} V_{\rm set}+\frac{2 {k_{\varv}}}{L_c}) \nonumber \\
      &-\frac{{K_p} V_{\rm set}^2 {\omega_1} (-\frac{{k_{\varv}^2} (\frac{{i_0}^2}{V_{\rm set}}+\frac{{i_{q0}}}{L_c {\omega_1}})}{V_{\rm set}}+\frac{{i_{q0}} L_c {\omega_1}}{V_{\rm set}}+1)}{L_c}>0.
      \label{Eq::StabilityConditionEMT_VirtualImpedance}
    \end{align}
    A useful expression from \eqref{Eq::StabilityConditionEMT_VirtualImpedance} can be obtained for $i_0$ and $i_{q0}$ equal to zero, which yields $K_p<\frac{2k_{\varv}(k_{\varv}^2+L_c^2\omega_1^2)}{L_c^2V_{\rm set}^2\omega_1}$. This shows that there is an upper bound for the maximum value of the frequency droop $K_p$, which is directly proportional to $k_{\varv}$. Summarizing the results from this section, we can make make the following observations: (i) the RMS model indicates that the system will remain small-signal stable for any value of the chosen damping coefficient $k_{\varv}$ and frequency droop $K_p$, under the condition that the reactive current injected by the VSC is equal to zero and (ii) without the virtual impedance controller, the EMT model indicates that the system is unstable for any positive value. 
    \vspace{-6pt}


\begin{figure}[t!]
    \centering
    \includegraphics[scale=0.26,trim={1cm 0.2cm 5.0cm 1.48cm}, clip]{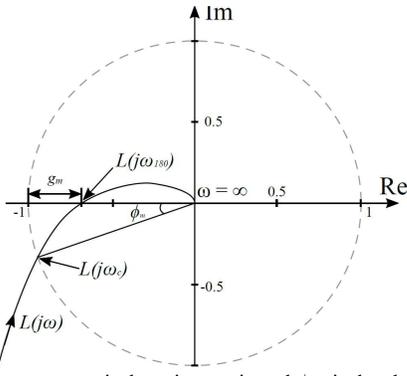}
    \vspace{-20pt}
    \caption{Nyquist curve: $g_m$ is the gain margin and $\phi_m$ is the phase \mbox{margin \cite{Skogestad2005}.}}
    \vspace{-15pt}
    \label{fig:nyquist}
\end{figure}

\section{Achievable performance of the RMS model - Frequency domain}
The traditional way to determine if the closed loop system is stable is to evaluate if all the roots of the closed loop characteristic polynomial lie in the left half plane. This approach is straightforward, however it gives little information about how to tune the gains in order to achieve system robustness. System robustness depends on the ability of the controller to damp poorly-damped modes. As presented in the previous section, the EMT model is able to capture poorly damped eigen-frequencies in power network, which the RMS cannot. These modes are the difference between the two models. As a consequence, we can expect a small mismatch between the two models, if these eigen-frequencies are well-damped. To this end, in this section: (i) we identify the differences in control design between the two models, using the phase and the gain margins (stability margins inspired by Nyquist's stability \mbox{criterion) \cite{Skogestad2005}} and (ii) propose a robust control design of the grid-forming converters that allows for sufficient damping of critical eigen-frequencies in power network.  
\vspace{-8pt}
\subsection{Control design using RMS and EMT models}
Consider the system in Fig. \ref{fig:OpenLoopSystem}, where $C(s)P(s)$ is the open loop transfer function of the system, with $C(s)$ the controller and $P(s)$ the plant. We seek to find a $C(s)$ that guarantees robust stability of the system. To this end, we use the Nyquist's criterion, which says that if the open loop transfer function encircles (-1, 0), then the closed loop control system is unstable \cite{Skogestad2005}. Moreover, we analyze the stability margin based on the gain margin ($g_m$) and phase margin $\phi_m$ which are relative measures of the distance of the Nyquist curve to the critical point (-1, 0), Fig. \ref{fig:nyquist}.

We introduce the open loop transfer function $L(s)$, which consists of the transfer functions $P(s)$ and $C(s)$, which describe the plant and controller dynamics, respectively.
\begin{equation}
    L(s) = P(s)C(s)
\end{equation}
where $C(s)$ = $\frac{K_p}{s}$, $P(s)$ = $G_{\theta P}^{\rm EMT}(s)$ for EMT modeling and $P(s)$ = $G_{\theta P}^{\rm RMS}(s)$ for RMS modeling.

The gain margin is calculated as follows: 
\begin{equation}
    g_m = - \frac{1}{|L(j\omega_{180})|}
    \label{Eq:gm}
\end{equation}
where $\omega_{180}$ is called phase crossover frequency and is the lowest frequency where the Nyquist plot intersects the negative real axis. The gain margin implies how much the controller gain can be increased before reaching the stability limit. Similarly, the phase margin is defined as:
\begin{equation}
    \phi_m = \pi + \arg L(j\omega_{c})
    \label{Eq::phasemargin}
\end{equation}
where $\omega_{c}$ is defined as the lowest frequency where the loop transfer function $L(s)$ has unit magnitude and implies the amount of phase lag required to reach the stability limit.
\subsubsection{RMS modeling - Open loop transfer function}
As in Section \ref{SectionIIIC} the assumption is that the high pass filter acts as pure resistance ($\omega_{\varv}$=0). To get better intuition about the stability properties, we also neglect the reactive current ($i_{q0}\approx0$). This assumption is reasonable under normal system operation and small-disturbances, since we want our generation units (PE devices in zero-inertia systems) to operate close to unity power factor at steady-state.
Hence, $L_{\rm RMS}(s)$ is an integrator transfer function. With respect to the stability margins, the characteristics of an integrator transfer function, are the infinite gain margin due to the non-defined phase crossover frequency $\omega_{180}$ and the $90^o$ degrees phase margin \cite{Skogestad2005}. As for the gain crossover frequency $\omega_{c}$ (see Fig. \ref{fig:nyquist}), it can be calculated as follows:  
\begin{equation}
    \centering
    \omega_c^{\rm RMS}=\frac{K_p L_c( V_{\rm set}^2-k_{\varv}^2 i_0^2)}{k_{\varv}^2+L_c^2}.
    \label{Eq::GainMarginRMS}
\end{equation}
Looking at the derived expression for $\omega_c$, the following observations can be made:
\begin{itemize}
    \item The higher the damping coefficient $k_{\varv}$ of the virtual impedance controller, the smaller the value of the gain crossover frequency becomes.
    \item Increasing the frequency droop $K_p$ leads to higher value of the gain crossover frequency.
    \item When the active current $i_{d0}$ ($i_{d0}=i_0$, since $i_{q0}\approx0$) of the VSC increases, the maximum obtainable gain crossover frequency decreases. 
\end{itemize}
\subsubsection{EMT modeling - Open loop transfer function}
Similar to the RMS modeling case, we neglect the reactive current and the cut-off frequency of the virtual impedance controller's filter. As it can be seen from \eqref{Eq:GpiEMT}, $G_{\theta P}^{\rm EMT}(s)$ has two zeros that, depending on the value of $i_0$, can either lie on the imaginary axis or one at the LHP and the other at the RHP.
As a result, we expect limitations on the control tuning in terms of system robustness and control performance. First, the phase crossover frequency is calculated as ($\Im{(L({j\omega_{180}}))}$=0):
\begin{equation}
    \centering
    \omega_{180}^{\rm EMT}=\frac{\sqrt{k_{\varv}^2+L_c^2 \omega_1^2}}{L_c}.
    \label{Eq::CrossoverEMT}
\end{equation}
Substituting $\omega_{180}^{\rm EMT}$ into \eqref{Eq:gm}, we can express $K_p$ as a function of the gain margin. The relation between $K_p$ and $g_{m}^{\rm EMT}$ is given by:
\begin{equation}
    \centering
    K_p^{\rm max} = \frac{2k_{\varv}(L_c^2+k_{\varv}^2)}{g_{m}^{\rm EMT}L_c^2(V_{\rm set}^2-i_{0}^2k_{\varv}^2)}.
    \label{Eq::droop contraint}
\end{equation}
As shown in \eqref{Eq::droop contraint}, the maximum obtainable value of the frequency droop is proportionally related to the damping coefficient $k_{\varv}$ of the virtual impedance control unit. The higher its value, the higher the maximum obtainable frequency droop. This relation holds for low active currents. 

In order to determine the gain crossover frequency, \eqref{Eq::phasemargin} is used with $C(s)=\frac{K_p}{s}$ and $P(s)=G_{\theta P}^{\rm EMT}(s)$. The Laplace complex variable $s$ is set to $j\omega_c^{\rm EMT}$ and $|L(j\omega_c^{\rm EMT})|=1$ is solved for $\omega_c^{\rm EMT}$. This yields the following polynomial: 
\begin{align}
     \centering
 f(x) &= -x^3 L_c^4+x^2 (2 L_c^4-2 k_{\varv}^2 L_c^2) \nonumber \\
 &-x (k_{\varv}^4+2 k_{\varv}^2 L_c^2+L_c^4)+K_p^2 L_c^2 (V_{\rm set}^2-k_{\varv}^2 i_0^2)^2 
\end{align}
where $x=(\omega_c^{\rm EMT})^2$. $f(x)$ is a cubic polynomial, thus it is possible to derive its roots analytically. Out of the three possible roots, we keep the one that is a positive real number and has the smallest magnitude, since $\omega_c$ is the lowest frequency where the phase of $L(s)$ is $-180^o$ \cite{Skogestad2005}. 
Since $\omega_c$ is the solution of $f(x)$ with the smallest magnitude, first order Taylor series expansion around 0 is utilized. This is similar to disregarding the second and third order term of $f(x)$. The analytical expression for the root of polynomial $f(x)$ is approximately given by:
\begin{equation}
    \centering
    \omega_c^{\rm EMT}\approx\frac{K_p L( V_{\rm set}^2-k_{\varv}^2 i_0^2)}{k_{\varv}^2+L_c^2}.
    \label{Eq::GainMarginEMT}
\end{equation}
It can be seen, that the gain crossover frequency with EMT modeling is equal to the one derived using the RMS model of the VSC in \eqref{Eq::GainMarginRMS}. The conclusions derived  in the case of RMS modeling, hold for EMT modeling as well.
Having calculated the gain crossover frequency, we can estimate the phase margin using \eqref{Eq::phasemargin}. A phase margin larger than $40^o$ is usually required to achieve robust performance and stability \cite{Holmes2009,Harnefors1998}. In our work, to increase robustness against system uncertainty, we impose as condition for the phase margin to be greater than $80^o$. As mentioned above, $K_p$ can be considered equal to $K_p^{\rm EMT}$=$\frac{2k_{\varv}(L^2+k_{\varv}^2)}{g_m^{\rm EMT}L^2(V_{\rm set}^2-i_{0}^2k_{\varv}^2)}$. Minimum phase margin of $80^o$ entails that $\arg({L(j\omega_c^{\rm EMT})|_{\footnotesize K_p=K_p^{\rm EMT}}})\geq -100^o$. This implies that $\Im({L(j\omega_c^{\rm EMT})|_{\footnotesize K_p=K_p^{\rm EMT}}}$) $\geq$ $\arctan(-100^o)$ $\Re({L(j\omega_c^{\rm EMT})|_{\footnotesize K_p=K_p^{\rm EMT}}})$. Thus, the condition that needs to be satisfied is:
\begin{equation}
\small
\centering
    k_{\varv} \geq \sqrt{-\frac{(g_m^{\rm EMT})^2 L_c^2}{(g_m^{\rm EMT})^2-24 g_m^{\rm EMT}-4}}
    \label{Eq::virtualImpedance_gainMargin}
\end{equation}
where $2<g_m^{\rm EMT}<24.2$. This means that, for a minimum phase margin of $80^o$, the gain margin can be 24.2 at maximum. For lower minimum phase margin, the maximum achievable gain margin decreases (e.g. if $\phi_m\geq45$, then $k_{\varv} \geq \sqrt{-\frac{{g_m^{\rm EMT}}^2 L_c^2}{(g_m^{\rm EMT})^2-4 g_m^{\rm EMT}-4}}$, where $0<g_m^{\rm EMT}<4.83$). This results in faster response of the active power controller, but also in decrease of system robustness (lower gain margin).
\subsubsection{Comparison of RMS and EMT model}
\begin{figure}[t]
    \centering
    \includegraphics[width=\linewidth,trim={3.3cm 11cm 3.2cm 11cm},clip]{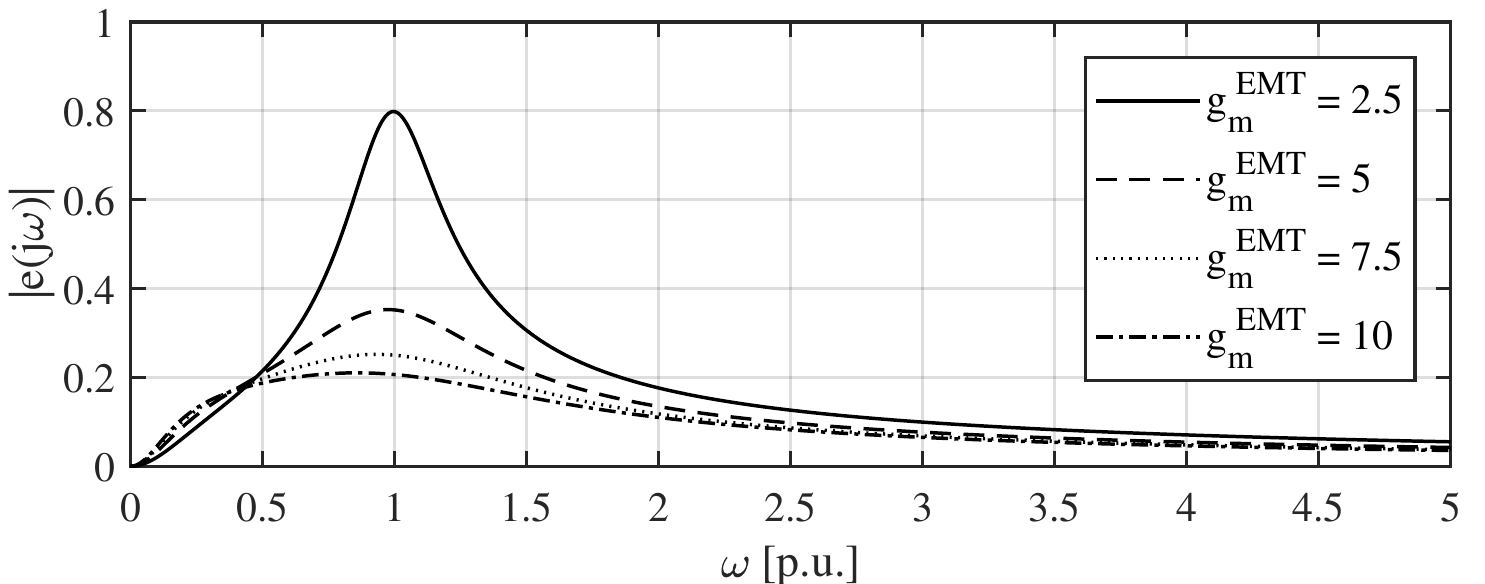}
    \vspace{-20pt}
    \caption{Singular values of $e(s)$ transfer function for different values of the gain margin.}
    \label{fig:bodeplotE(s)}
        \vspace{-15pt}
\end{figure}

Having derived the expressions for calculating the control gains of the frequency droop and virtual impedance controllers, we analyze the transfer function \mbox{$e(s)$=$G_{Pcl}^{\rm EMT}(s) - G_{Pcl}^{\rm RMS}(s)$}, which represents the estimation error of the RMS model in respect to the EMT model. 
Fig. \ref{fig:bodeplotE(s)} depicts the bodeplot (singular values) of the transfer function $e(s)$. As it can be seen in the figure, for small values of $g_m^{\rm EMT}$, the magnitude of $e(s)$ is of significance around the nominal frequency of the system. This implies that the power signal estimated using the RMS model will present an important error with respect to the actual power signal (derived using the EMT model). Increasing the gain margin decreases the estimation error, as shown in the same figure. Bodeplot of $e(s)$ is used to measure the accuracy of the RMS model. 

Moreover, as shown in the figure, the singular values $e(s)$ at steady state, $\omega=0$ are equal to zero, which was expected since the RMS model is able to preserve the dc-gain of the EMT state-space model. It is also depicted that during a dynamic event $\omega>0$, the RMS model presents a mismatch with respect to the EMT model.  

To summarize the analysis and the results from this section:
\begin{itemize}
    \item Due to the non-defined phase crossover frequency, the RMS model has an infinite gain margin, which gives a wrong indication about the robustness of the system. That implies that the RMS model will indicate that the system will be small-signal stable for any value of the chosen $k_{\varv}$ and $K_p$. This conclusion also agrees with the sufficient conditions \eqref{Eq::Pole1} derived in the previous section. 
    \item In the case of EMT modeling, the phase crossover frequency is well defined and, hence, the gain margin has a finite value. High values of $g_m^{\rm EMT}$ and $\phi_m$ infers that high frequency modes associated with the network dynamics are well damped. This results in smaller mismatch between the RMS and EMT models (see Fig. \ref{fig:bodeplotE(s)}), since the dominant modes are of lower frequency. It was also shown that the combination of the chosen damping coefficient $k_{\varv}$ and frequency droop $K_p$ impact the suitability of the RMS modeling of zero-inertia systems. 
\end{itemize}

\vspace{-10pt}
\section{Validation of stability analysis - Test Case}

\begin{figure}[t]
    \centering
    \includegraphics[scale=0.15,trim={-0.8cm 0.0cm 0.6cm -0.2cm}, clip]{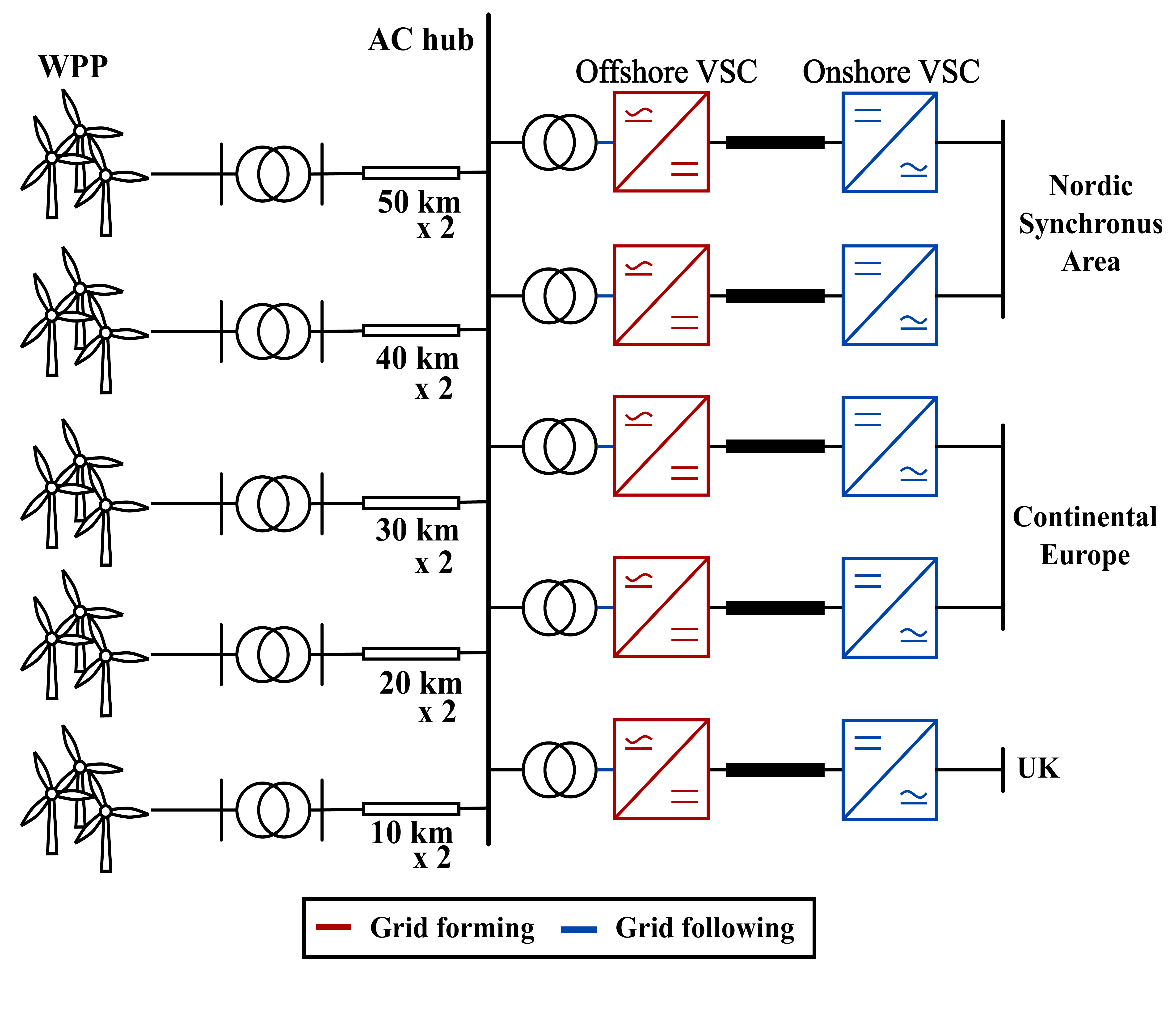}
              \vspace{-15pt}
    \caption{Test system of case study.}
    \label{fig:SystemFig}
    \vspace{-15pt}
\end{figure}

In this section, we present a case study where we highlight the appropriateness of the RMS model for simulating power disturbances. Our goal is to compare the active power response, obtained by the RMS and EMT models, for different values of the gain margin $g_m^{\rm EMT}$. To show this, we consider three cases, where we use \eqref{Eq::droop contraint} and \eqref{Eq::virtualImpedance_gainMargin} in order to tune the damping coefficient $k_{\varv}$ and the frequency droop $K_p$. The control parameters have the following values: 
\begin{itemize}
    \item Case 1: $g_m^{\rm EMT}$=2.5, $k_{\varv}$=0.0658 and $K_p$ = 0.0584.
    \item Case 2: $g_m^{\rm EMT}$=10, $k_{\varv}$=0.1667 and $K_p$ = 0.0569.
    \item Case 3: $g_m^{\rm EMT}\approx0$, $k_{\varv}$=0 and $K_p$ = 0.01.
\end{itemize}
It should be mentioned that all of the grid-forming converters are homogeneous and tuned similarly. Moreover, in all three cases, we monitor the active power response of a grid-forming converter for a step change of its power reference signal $P_{\rm ref}$.

The system depicted in Fig. \ref{fig:SystemFig} is used for our analysis and comparison between RMS and EMT modeling. The system represents an AC configuration for integrating large offshore wind power. The wind power generated by the wind farms is transferred through High-Voltage-Alternating-Current (HVAC) cables (220 kV) to an Offshore Energy Hub, and from there it can be distributed to different countries as shown in Fig. \ref{fig:SystemFig}. The cable parameters are taken from manufacturer data sheets \cite{ABB:cable}. Five point-to-point High Voltage Direct Current (HVDC) links are used to transfer the amount of power to the interconnected onshore grids. The onshore grid is represented by a grid equivalent with inertial and primary response \cite{Misyris2019,Tilman2017}. The converters on the offshore side operate in grid-forming mode, where they set the voltage and the frequency of the offshore system. Additionally, each converter on the onshore side operates in grid-following mode, providing constant reactive power and regulating the voltage of the HVDC-link \cite{Thams2017}. A generic model of a wind farm (WPP) is used based on a dynamic equivalent presented in \cite{Chaspierre2017DynamicEO}. As base power, we consider Sb = 1000 MVA. We consider that the rated power of all offshore and onshore converters is 1100 MVA. As for the wind farms, their rated power is \mbox{700 MVA}. 
\vspace{-5pt}
\subsection{Active power response with virtual impedance}
\begin{figure}
    \centering
    \includegraphics[width=\linewidth,trim={3.3cm 11cm 3.2cm 11cm},clip]{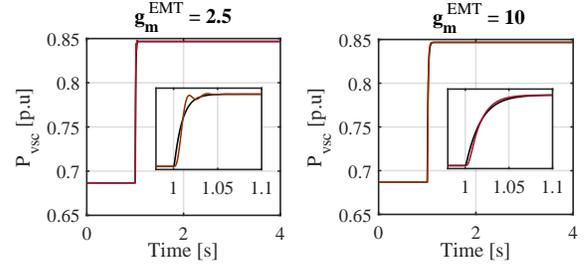}
        \vspace{-20pt}
    \caption{Comparison between RMS (black solid line) and EMT (red solid line) for a step change of $P_{\rm ref}$ for one of the offshore VSCs.}
    \label{fig:StepChange}
        \vspace{-7pt}
\end{figure}

Fig. \ref{fig:StepChange} depicts the active power absorbed by the VSC. A 0.2 p.u. step change of $P_{\rm ref}$ is performed. As shown in the figure, in the case of lower gain margin, the mismatch between the RMS and EMT is larger compared to the one with higher gain margin. The reason for that is that in case of lower gain margin, the damping coefficient $k_{\varv}$ is lower \eqref{Eq::virtualImpedance_gainMargin}. This confirms the results in Fig. \ref{fig:bodeplotE(s)}. Due to the limited damping provided by the virtual impedance controller, electro-magnetic oscillations appear in the power signal, which increases the mismatch between RMS and EMT (since the RMS is unable to capture high frequency modes). 
Moreover, one could notice that the time response of the system is much faster in the case of lower $g_m^{\rm EMT}$, but it is similar for both the RMS and EMT models. This comes to verify the condition derived for the gain crossover frequency (indication of the closed loop bandwidth) in the previous section, where it was shown that the system response becomes slower when the damping coefficient $k_{\varv}$ increases in \eqref{Eq::GainMarginRMS} and \eqref{Eq::GainMarginEMT}. Overall, the RMS model predicts that the system is stable, however it underestimates the effect of network dynamics on the system response, in terms of quality of the signal.
\vspace{-5pt}
\subsection{Active power response without virtual impedance}
\begin{figure}
    \centering
    \includegraphics[width=\linewidth,trim={3.3cm 11cm 3.2cm 11cm},clip]{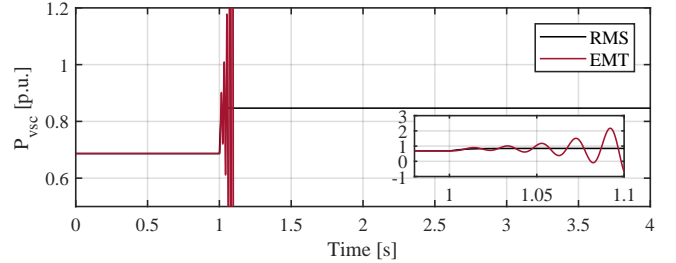}
        \vspace{-20pt}
    \caption{Comparison between RMS (black solid line) and EMT (red solid line) for a step change of $P_{\rm ref}$ for one of the offshore VSCs when neglecting the virtual impedance controller ($g_m^{\rm EMT}\approx$0).}
    \vspace{-14pt}
    \label{fig:StepChangeUnstable}
\end{figure}

Fig. \ref{fig:StepChangeUnstable} presents a comparison between the RMS and EMT model, when the virtual impedance controller is neglected, with $g_m^{\rm EMT}\approx$ 0 ($k_{\varv}=0$) and $K_p=0.01$. As it can be seen in the figure, even though the EMT model indicates that the system is unstable, the RMS predicts that the system is stable. The EMT model indicates that there is an unstable mode at the nominal frequency, since the time period of the oscillations is equal to 20 $ms$. This verifies the results presented in Fig. \ref{fig:ComparisonAnalyticalWithout} and the sufficient conditions derived using the Routh-Hurwitz stability criterion. The results show the important role of the virtual impedance controller when simulating zero- and low- inertia systems. When the RMS model is used, neglecting the virtual impedance can lead to overestimation of the stability boundaries of the system.


\vspace{-5pt}
\section{Conclusions}
In this paper, we studied the appropriateness of phasor-approximation models when simulating zero-inertia systems.
Sufficient conditions were derived for the system stability using Routh-Hurwitz stability criterion. Then, the gain and phase margins of the open-loop system were derived, in order to evaluate the performance and robustness of the grid-forming VSC for both RMS and EMT modeling. 
The main conclusions of the linearized and time-domain simulation analyses yield the following:
\begin{itemize}
    \item RMS modeling of grid-forming converter can lead to the identification of wrong stability boundaries. Based on \eqref{Eq::Pole1}, the system will always be small-signal stable for any positive value of $K_p$, which in reality does not hold. This was also verified with time domain simulations. It should be mentioned that the only way for the system to become unstable, when RMS modeling is used, is to have a significant drop of the terminal voltage of the converter. 
    \item It was shown that the mismatch between the RMS and EMT modeling is indirectly proportional to the gain margin. The higher the damping of modes corresponding to electro-magnetic dynamics (that cannot be captured by phasor-models) the smaller the mismatch between the two models. Under this condition (appropriate damping of modes corresponding to electro-magnetic dynamics), RMS modeling is appropriate for simulating power disturbances.
    \item Our findings suggest that system operators could keep on using the phasor-approximation model in the presence of the zero-inertia systems for monitoring the active power sharing between grid-forming converters as long as the eigen-frequencies in the power network are well damped. 
\end{itemize}

Future work will focus on investigating the impact of high reactive currents injected by the VSC, which might occur during large voltage deviations. Moreover, a categorization of different types of small and large disturbances is needed, in order to understand under which conditions we need to use EMT modeling. 
\vspace{-5pt}
\appendix
\numberwithin{equation}{section}
\setcounter{equation}{0}
We derive the transfer function describing the relation between $\Delta \theta$ and $\Delta \boldsymbol{i}$. We consider that $\boldsymbol{\varv} = V_{\rm set}$ and $\boldsymbol{\varv_{g}}^c=V_g$ at steady state. Neglecting the resistance of the phase reactor and using perturbation quantities of $\theta$, $\boldsymbol{\varv}$ and $\boldsymbol{i}$, \eqref{Eq:phasereactor} is expressed as:
\begin{equation}
    V_{\rm set}+\Delta \boldsymbol{\varv}-[s+j(\omega_1+\dot{\Delta}\theta)]L_c(\boldsymbol{i_0}+\Delta \boldsymbol{i})=V_ge^{-j(\theta_0+\Delta\theta)}.
    \label{Eq:ap1}
\end{equation}
Since no $QV$ droop controller is considered, \mbox{$\Delta \boldsymbol{\varv} = - H_{\rm HP}(s)\Delta \boldsymbol{i}$}. Moreover, we can neglect the term $\Delta \theta$ since for normal operation $\Delta \theta \approx 0$. As shown in the stability analysis, large value of frequency droops cause small-signal instability. Thus, small values of frequency droops are appropriate for guaranteeing robust small-signal stability (high damping ratio of the active power controller). Small values of frequency droops lead to small frequency deviation. Thus, neglecting $\dot{\Delta} \theta$ is reasonable. Additionally, in case of RMS modeling the laplace variable s is equal to zero, since the network dynamics are disregarded. Last but not least, the term $e^{-\Delta\theta}\approx1-j\Delta\theta$. The approximation stands for small variation of $\Delta\theta\approx0$, since $\cos{(\Delta\theta)}\approx0$ and $\sin{(\Delta\theta)}\approx\Delta\theta$. This can be explained by:
\begin{equation}
    \centering
    e^{-j\Delta\theta} = \underbrace{\cos{(\Delta\theta)}}_1-\underbrace{j\sin{(\Delta\theta)}}_{j\Delta\theta}.
    \label{Eq:a}
\end{equation}
After these simplifications \eqref{Eq:ap1} is expressed as:
\begin{equation}
    \begin{split}
    [H_{\rm HP}(s)+j\omega_1L_c]\Delta \boldsymbol{i}=&j(V_ge^{-j\theta_0})\Delta\theta \\
    & \underbrace{+ V_{\rm set}-j\omega_1L_c\boldsymbol{i_0}-V_ge^{-j\theta_0}}_{=0}
    \label{Eq:app2}
\end{split}
\end{equation}
where $V_{\rm set}-j\omega_1L_c\boldsymbol{i_0}-V_ge^{-j\theta_0}=0$ (steady state condition). Using \eqref{Eq:app2}, we can derive the relation between $\Delta\theta$ and $\Delta \boldsymbol{i}$:
\begin{equation}
    \Delta \boldsymbol{i}=\underbrace{\frac{j(V_{\rm set}-j\omega_1)L_c\boldsymbol{i_0}}{H_{\rm HP}(s)+j\omega_1L_c}}_{G_{\theta\Delta i}(s)}\Delta\theta.
    \label{Eq:DiDt}
\end{equation}
Similar to \cite{Harnefors2018}, by using perturbation variables we can express the active power variation of the VSC as:
\begin{equation}
    \centering
    \Delta P = \Re{(V_{\rm set}\Delta \boldsymbol{i^*}+\boldsymbol{i_0^*}{\Delta \boldsymbol{\varv}})}.
    \label{Eq:app3}
\end{equation}
Substituting \eqref{Eq:DiDt} into \eqref{Eq:app3} and considering that \mbox{$\Delta \boldsymbol{\varv} = - H_{\rm HP}(s)\Delta i$} yields:
\begin{equation}
\centering
    \Delta P = \underbrace{\Re{(V_{\rm set}G_{\theta\Delta i}^*(s)-\boldsymbol{i_0^*}H_{\rm HP}(s)G_{\theta\Delta i}(s))}}_{G_{\theta P}^{\rm RMS}(s)} \Delta\theta.
\end{equation}

\vspace{-10pt}

\bibliographystyle{IEEEtran}
\bibliography{IEEEabrv,References}

\begin{thebibliography}{10}
\providecommand{\url}[1]{#1}
\csname url@samestyle\endcsname
\providecommand{\newblock}{\relax}
\providecommand{\bibinfo}[2]{#2}
\providecommand{\BIBentrySTDinterwordspacing}{\spaceskip=0pt\relax}
\providecommand{\BIBentryALTinterwordstretchfactor}{4}
\providecommand{\BIBentryALTinterwordspacing}{\spaceskip=\fontdimen2\font plus
\BIBentryALTinterwordstretchfactor\fontdimen3\font minus
  \fontdimen4\font\relax}
\providecommand{\BIBforeignlanguage}[2]{{%
\expandafter\ifx\csname l@#1\endcsname\relax
\typeout{** WARNING: IEEEtran.bst: No hyphenation pattern has been}%
\typeout{** loaded for the language `#1'. Using the pattern for}%
\typeout{** the default language instead.}%
\else
\language=\csname l@#1\endcsname
\fi
#2}}
\providecommand{\BIBdecl}{\relax}
\BIBdecl

\bibitem{Milano2018}
F.~Milano, F.~D{\"o}rfler, G.~Hug, D.~J. Hill, and G.~Verbi{\v{c}},
  ``Foundations and challenges of low-inertia systems (invited paper),'' in
  \emph{2018 Power Systems Computation Conference (PSCC)}, Jun. 2018, pp.
  1--25.

\bibitem{markovic2019}
\BIBentryALTinterwordspacing
U.~Markovic, O.~Stanojev, E.~Vrettos, P.~Aristidou, and G.~Hug, ``Understanding
  stability of low-inertia systems,'' Feb. 2019. [Online]. Available:
  \url{engrxiv.org/jwzrq}
\BIBentrySTDinterwordspacing

\bibitem{Collados2019}
C.~{Collados-Rodriguez}, M.~{Cheah-Mane}, E.~{Prieto-Araujo}, and
  O.~{Gomis-Bellmunt}, ``Stability analysis of systems with high vsc
  penetration: Where is the limit?'' \emph{IEEE Transactions on Power
  Delivery}, pp. 1--1, 2019.

\bibitem{Misyris2018}
G.~S. {Misyris}, S.~{Chatzivasileiadis}, and T.~{Weckesser}, ``Robust frequency
  control for varying inertia power systems,'' in \emph{2018 IEEE PES
  Innovative Smart Grid Technologies Conference Europe (ISGT-Europe)}, Oct.
  2018, pp. 1--6.

\bibitem{Misyris2019}
G.~S. {Misyris}, J.~A. {Mermet-Guyennet}, S.~{Chatzivasileiadis}, and
  T.~{Weckesser}, ``Grid supporting vscs in power systems with varying inertia
  and short-circuit capacity,'' in \emph{2019 IEEE Milan PowerTech}, Jun. 2019,
  pp. 1--6.

\bibitem{Farrokhabadi2019}
M.~{Farrokhabadi} \emph{et~al.}, ``Microgrid stability definitions, analysis,
  and examples,'' \emph{IEEE Transactions on Power Systems}, vol.~35, no.~1,
  pp. 13--29, Jan 2020.

\bibitem{kundur}
P.~Kundur, N.~J. Balu, and M.~G. Lauby, \emph{Power system stability and
  control}.\hskip 1em plus 0.5em minus 0.4em\relax McGraw-hill New York, 1994,
  vol.~7.

\bibitem{Turhan2008}
T.~Demiray, ``\BIBforeignlanguage{en}{Simulation of power system dynamics using
  dynamic phasor models},'' Ph.D. dissertation, ETH Zurich, 2008.

\bibitem{Vorobev2018}
P.~{Vorobev}, P.~{Huang}, M.~{Al Hosani}, J.~L. {Kirtley}, and K.~{Turitsyn},
  ``High-fidelity model order reduction for microgrids stability assessment,''
  \emph{IEEE Trans. Power Syst.}, vol.~33, no.~1, pp. 874--887, Jan. 2018.

\bibitem{Gu2018}
Y.~{Gu}, N.~{Bottrell}, and T.~C. {Green}, ``Reduced-order models for
  representing converters in power system studies,'' \emph{IEEE Transactions on
  Power Electronics}, vol.~33, no.~4, pp. 3644--3654, Apr 2018.

\bibitem{Purba2019}
V.~{Purba}, B.~B. {Johnson}, S.~{Jafarpour}, F.~{Bullo}, and S.~V. {Dhople},
  ``Dynamic aggregation of grid-tied three-phase inverters,'' \emph{IEEE Trans.
  Power Syst.}, pp. 1--1, 2019.

\bibitem{Purba2019b}
V.~{Purba}, B.~B. {Johnson}, M.~{Rodriguez}, S.~{Jafarpour}, F.~{Bullo}, and
  S.~V. {Dhople}, ``Reduced-order aggregate model for parallel-connected
  single-phase inverters,'' \emph{IEEE Trans. Energy Convers.}, vol.~34, no.~2,
  pp. 824--837, Jun. 2019.

\bibitem{Rocabert2012}
J.~{Rocabert}, A.~{Luna}, F.~{Blaabjerg}, and P.~{Rodríguez}, ``Control of
  power converters in ac microgrids,'' \emph{IEEE Trans. Power Electron.},
  vol.~27, no.~11, pp. 4734--4749, Nov. 2012.

\bibitem{Kundur2003}
P.~{Kundur} \emph{et~al.}, ``Definition and classification of power system
  stability ieee/cigre joint task force on stability terms and definitions,''
  \emph{IEEE Trans. Power Syst.}, vol.~19, no.~3, pp. 1387--1401, Aug 2004.

\bibitem{Raza2018}
M.~{Raza}, E.~{Prieto-Araujo}, and O.~{Gomis-Bellmunt}, ``Small-signal
  stability analysis of offshore ac network having multiple vsc-hvdc systems,''
  \emph{IEEE Trans. Power Del.}, vol.~33, no.~2, pp. 830--839, Apr 2018.

\bibitem{Pogaku2007}
N.~{Pogaku}, M.~{Prodanovic}, and T.~C. {Green}, ``Modeling, analysis and
  testing of autonomous operation of an inverter-based microgrid,'' \emph{IEEE
  Trans. Power Electron.}, vol.~22, no.~2, pp. 613--625, Mar. 2007.

\bibitem{SIMPSONPORCO20132603}
J.~W. Simpson-Porco, F.~Dörfler, and F.~Bullo, ``Synchronization and power
  sharing for droop-controlled inverters in islanded microgrids,''
  \emph{Automatica}, vol.~49, no.~9, pp. 2603 -- 2611, 2013.

\bibitem{Papangelis2018}
L.~{Papangelis}, M.~{Debry}, T.~{Prevost}, P.~{Panciatici}, and T.~{Van
  Cutsem}, ``Stability of a voltage source converter subject to decrease of
  short-circuit capacity: A case study,'' in \emph{2018 Power Systems
  Computation Conference (PSCC)}, June 2018, pp. 1--7.

\bibitem{Harnefors2018}
L.~{Harnefors}, M.~{Hinkkanen}, U.~{Riaz}, F.~M.~M. {Rahman}, and L.~{Zhang},
  ``Robust analytic design of power-synchronization control,'' \emph{IEEE
  Trans. Ind. Electron}, vol.~66, no.~8, pp. 5810--5819, Aug. 2019.

\bibitem{ZhangThesis}
L.~Zhang, ``Modeling and control of vsc-hvdc links connected to weak ac
  systems,'' Ph.D. dissertation, KTH, Electrical Machines and Power
  Electronics, 2010, qC20100607.

\bibitem{Zhang2010}
L.~{Zhang}, L.~{Harnefors}, and H.~{Nee}, ``Power-synchronization control of
  grid-connected voltage-source converters,'' \emph{IEEE Trans. Power Syst.},
  vol.~25, no.~2, pp. 809--820, May 2010.

\bibitem{Skogestad2005}
S.~Skogestad and I.~Postlethwaite, \emph{Multivariable Feedback Control:
  Analysis and Design}.\hskip 1em plus 0.5em minus 0.4em\relax John Wiley \&
  Sons, 2005.

\bibitem{rahman2002analytic}
Q.~I. Rahman, G.~Schmeisser \emph{et~al.}, \emph{Analytic theory of
  polynomials}.\hskip 1em plus 0.5em minus 0.4em\relax Oxford University Press,
  2002, no.~26.

\bibitem{Holmes2009}
D.~G. {Holmes}, T.~A. {Lipo}, B.~P. {McGrath}, and W.~Y. {Kong}, ``Optimized
  design of stationary frame three phase ac current regulators,'' \emph{IEEE
  Trans. Power Electron.}, vol.~24, no.~11, pp. 2417--2426, Nov. 2009.

\bibitem{Harnefors1998}
L.~{Harnefors} and H.~. {Nee}, ``Model-based current control of ac machines
  using the internal model control method,'' \emph{IEEE Trans. Ind Appl.},
  vol.~34, no.~1, pp. 133--141, Jan. 1998.

\bibitem{ABB:cable}
ABB. (2010, 04) {XLPE} submarince cable systems: Attachment to {XLPE} land
  cable systems. Rev. 5.

\bibitem{Tilman2017}
T.~{Weckesser} and T.~{Van Cutsem}, ``\BIBforeignlanguage{English}{Equivalent
  to represent inertial and primary frequency control effects of an external
  system},'' \emph{\BIBforeignlanguage{English}{IET Gener., Transm. Dis.}},
  vol.~11, pp. 3467--3474(7), Sep. 2017.

\bibitem{Thams2017}
F.~{Thams}, R.~{Eriksson}, and M.~{Molinas}, ``Interaction of droop control
  structures and its inherent effect on the power transfer limits in
  multiterminal vsc-hvdc,'' \emph{IEEE Trans. Power Del.}, vol.~32, no.~1, pp.
  182--192, Feb. 2017.

\bibitem{Chaspierre2017DynamicEO}
G.~{Chaspierre}, P.~{Panciatici}, and T.~{Van Cutsem}, ``Dynamic equivalent of
  a distribution grid hosting dispersed photovoltaic units,'' in \emph{Proc. of
  IREP'2017 Symp.}, 2017.

\end{thebibliography}

\vspace{-10pt}

\end{document}